\documentclass[fullpaper]{jpsj3}
\usepackage{txfonts}

\usepackage{siunitx}
\usepackage{textcomp}

\usepackage{algpseudocode}
\usepackage{algorithm}

\usepackage{here}

\usepackage{tabularx}

\usepackage{graphicx}
\usepackage{float}
\usepackage{textcomp}
\usepackage{amsmath}
\usepackage{listings}
\usepackage{url}

\usepackage{tikz}
\usetikzlibrary{trees}
\usepackage{bm}
\usepackage[whole]{bxcjkjatype}

\title{Efficient Construction of Feasible Solutions in Column Generation \\using Quantum Annealing}

\author{Taisei Takabayashi$^{1, 2}$, Naoki Maruyama$^{1, 2*}$, Takuma Yoshihara$^{1, 3}$, \\ Renichiro Haba$^{1, 2}$, and Masayuki Ohzeki$^{1, 2, 4}$}
\inst{
Sigma-i Co., Ltd., Tokyo, 108-0075, Japan$^{1}$ \\
Graduate School of Information Sciences, Tohoku University, Miyagi 980-8564, Japan$^{2}$ \\
Department of Engineering, Tokyo Denki University, Tokyo 120-8551, Japan$^{3}$ \\
Department of Physics, Tokyo Institute of Technology, Tokyo, 152-8551, Japan$^{4}$\\
} 

\abst{
Column generation (CG) has been used to solve constrained 0-1 quadratic programming problems. The pricing problem, which is iteratively solved in CG, can be reduced to an unconstrained 0-1 quadratic programming problem, allowing for the efficient application of quantum annealing (QA). The solutions obtained by CG are continuous relaxations, which cannot be practically used as feasible 0-1 solutions. In this paper, we propose a postprocessing method for constructing feasible 0-1 solutions from the continuous relaxations obtained through CG. The proposed technique consists of two phases: (i) mapping the continuous CG solution to a feasible 0-1 solution and (ii) applying a constraint-aware local search to improve that solution's quality. Numerical experiments on randomly generated problems demonstrate that CG with the proposed postprocessing yields solutions comparable to commercial solvers with significantly reduced computation time. Consequently, the postprocessing enables CG with QA to obtain high-quality approximate solutions faster.
}

\begin{document}
\maketitle
\renewcommand{\algorithmicrequire}{\textbf{Input:}}
\renewcommand{\algorithmicensure}{\textbf{Output:}}

%%%%%%%%%%%%%%%%%%%%%%%%%%%%%%%%%%%%%%%%%%%%%%%%%%%%%%%%%%%%%%%%%%%%%%%%%%%%%%%%
\section{Introduction}
Quantum annealing (QA) is a generic solver for combinatorial optimization problems\cite{kadowaki1998quantum}. Since the development of the quantum annealer by D-Wave Systems, the application of QA to various fields has been studied\cite{rosenberg2016solving, venturelli2019reverse, neukart2017traffic, hussain_optimal_2020, doi:10.7566/JPSJ.94.024001, Haba2025, Ohzeki2019, haba_travel_2022, KoukiYonaga2022ISIJINT-2022-019, 10.3389/fcomp.2023.1285962, ide2020maximum, Arai2021code, Yamamoto2020, maruyama2021graph, amin2018quantum, o2018nonnegative, Sato2021, doi:10.7566/JPSJ.91.074008, Arai2021, hasegawa2023, doi:10.7566/JPSJ.94.034002, haba2025relaxationassistedreverseannealingnonnegativebinary}.

Current quantum annealers struggle with constrained optimization, as they handle only quadratic unconstrained binary optimization (QUBO) problems\cite{glover_quantum_2022}. Constraints are often encoded using penalty methods\cite{lucas2014ising}, adding many quadratic terms.  To address this, methods that relax equality constraints without the penalty method have been proposed\cite{ohzeki2020}.

In addition, inequality-constrained optimization problems require transforming inequalities using slack variables, necessitating additional binary variables. This reduces the size of problems that quantum annealers with limited qubits can address. In contrast, methods that utilize QA iteratively, such as Lagrangian relaxation\cite{Karimi2017, 53gd-2374, doi:10.7566/JPSJ.94.054003} and extended Lagrangian methods\cite{Yonaga2020inequality, https://doi.org/10.1002/qute.202300104, cellini_qal-bp_2024}, have been proposed to express QUBO without using slack variables. 

Among these iterative methods addressing inequality constraints, the method proposed by Hirama\cite{Hirama2023}, which applies QA to inequality-constrained optimization problems is focused on in this study. This approach is based on column generation (CG) to solve the continuous relaxation of the original problem through Dantzig--Wolfe decomposition\cite{f89fdec7-edd6-39e9-bb25-bc143df4e941}. CG involves alternating between solving the dual problem of the restricted master problem and solving the pricing problem using the dual solution. Since the pricing problem reduces to a QUBO problem\cite{bettiol:tel-03227417}, QA is utilized as an efficient method for obtaining approximate solutions to this problem. Moreover, methods combining CG and QA have been specifically proposed for certain problems\cite{Ossorio-Castillo2022, 10736594, wagner2024quantumsubroutinesbranchpriceandcutvehicle, PhysRevA.107.032426, 10313804}.

However, the approach in the literature\cite{Hirama2023} only provides continuous relaxation solutions, which cannot be directly used for the original 0-1 problem. To bridge this practical limitation, in this paper, we propose a postprocessing method to construct feasible 0-1 solutions from the solutions obtained by CG. The proposed postprocessing method consists of constructing feasible solutions from infeasible 0-1 solutions and local search. Numerical experiments on random problems demonstrate that the combination of CG and the proposed postprocessing method achieves approximate solutions comparable to those obtained by commercial general-purpose solvers, such as Gurobi, at significantly higher speeds as the problem size increases. From these results, it was demonstrated that feasible 0–1 solutions can be obtained by combining CG with QA and the proposed postprocessing method, and that this approach can serve as a fast approximate solver for large-scale problems.

%%%%%%%%%%%%%%%%%%%%%%%%%%%%%%%%%%%%%%%%%%%%%%%%%%%%%%%%%%%%%%%%
\section{Background}
In this paper, we address constrained quadratic programming problems of the following form:
\begin{equation}
\begin{aligned}
    \min_{\bm{x}} 
        & \quad \sum_{i j} Q_{i j} x_i x_j, \\
    \text{s.t.} 
        & \quad \sum_{i j} A_{k i j} x_i x_j \leq b_k, \quad \forall k\in\{1, \dots, m\}, \\
        & \quad x_i \in\{0, 1\}, \quad \forall i\in\{1, \dots, n\},
\label{constrained quadratic programming}
\end{aligned}
\end{equation}
where $Q$ and $A_k$ are upper triangular matrices of size $(n\times n)$, $b$ is a vector of size $m$, and $m$ represents the number of constraints. This problem is referred to as the ``original problem'' in this paper. To reduce computational complexity, the problem \eqref{constrained quadratic programming} can be relaxed through Dantzig--Wolfe decomposition\cite{f89fdec7-edd6-39e9-bb25-bc143df4e941} into a restricted master problem (RMP), formulated as follows:
\begin{equation}
\begin{aligned}
    \min _{\bm{\lambda}} 
        & \quad \sum_{p \in \overline{\mathcal{P}}} \sum_{i j} Q_{i j} x_i^p x_j^p \lambda^p, \\
    \text{s.t.} 
        & \quad \sum_{p \in \overline{\mathcal{P}}} \sum_{i j} A_{k i j} x_i^p x_j^p \lambda^p \leq b_k, \quad \forall k\in\{1, \dots, m\}, \\
        & \quad \sum_{p \in \overline{\mathcal{P}}} \lambda^p = 1, \\
        & \quad \lambda^p \geq 0, \quad \forall p \in \overline{\mathcal{P}}.
\label{restricted master problem}
\end{aligned}
\end{equation}
Here, $\overline{\mathcal{P}}=\{{\bm x}^1, \cdots, {\bm x}^p, \cdots \}$ is the set of extreme points, and each extreme point ${\bm x}^p$ corresponds to a solution of the original problem \eqref{constrained quadratic programming}. 
CG proposed in the literature\cite{bettiol:tel-03227417} iteratively constructs a manageable subset $\overline{\mathcal{P}}$. This involves solving a sequence of dual and pricing problems. 
If the objective value of the pricing problem with solution ${\bm x}^{\ast}$ is negative, the solution ${\bm x}^{\ast}$ is added to $\overline{\mathcal{P}}$, and the dual problem is solved repeatedly. Otherwise, the CG process terminates, and RMP \eqref{restricted master problem} is solved over the final $\overline{\mathcal{P}}$. Consequently, we can obtain the following continuous relaxation solution:
\begin{equation}
    X=\sum_{p\in\overline{\mathcal{P}}}\lambda_p {\bm x}^p \left({\bm x}^p\right)^{\rm T}.
    \label{X}
\end{equation}
Since the pricing problem is reduced to a QUBO problem, Hirama's method applies QA to solve this\cite{Hirama2023}. Although obtaining an exact solution within a practical computation time becomes challenging as the problem size $n$ increases, QA can be utilized as a method to obtain good approximate solutions for QUBO problems within a relatively short computation time.

However, the solution obtained through CG \eqref{X} is the continuous relaxation solution of the original problem \eqref{constrained quadratic programming}. Even in the prior research\cite{Hirama2023}, there is no mention of a method of constructing the 0-1 solution to the original problem \eqref{constrained quadratic programming}.

The solutions obtained via CG can also be employed in conjunction with exact algorithms, such as the branch-and-price method. In this method, the objective function value of the continuous relaxation from CG serves as a lower bound. Previous studies have applied the branch-and-price method\cite{210414dc-2516-3950-a574-223450def2d4} to specific problems like capacitated vehicle routing problem\cite{wagner2024quantumsubroutinesbranchpriceandcutvehicle}, using QA to solve the pricing problems. However, exact methods such as this require repeated CG running, resulting in significant computation time.

To address this issue, we propose a postprocessing method to swiftly derive feasible 0-1 solutions from CG's continuous relaxation results. We round the continuous relaxation $X$ to obtain an initial binary solution ${\bm x}_{\rm init}\in \{0,1\}^n$, which is refined via local search to ensure feasibility.

\section{Method}
The proposed postprocessing method transforms the continuous relaxation solutions obtained by CG \eqref{X} into feasible binary solutions. The method comprises two key processes: feasibility restoration and local optimization. To guide these processes, a measure called "efficiency" is defined for each variable. In our methods, efficiency evaluates the impact of flipping on both the objective function $p_i$ and the constraint satisfaction $w_{ik}$ (for each variable $x_i$ and constraint $k$). These are defined as follows: 
\begin{equation}
    p_i = f_i\left(Q_{ii}+\sum_{j=1}^{i-1}Q_{ji}x_j + \sum_{j=i+1}^{N}Q_{ij}x_j\right),
\label{p_i}
\end{equation}
\begin{equation}
    w_{ik} = f_i\left(A_{kii}+\sum_{j=1}^{i-1}A_{kji}x_j + \sum_{j=i+1}^{N}A_{kij}x_j\right),
\label{w_i}
\end{equation}
where $f_i$ represents the flip direction ($+1$ for flipping from $0$ to $1$, $-1$ for flipping from $1$ to $0$).
\begin{equation}
  f_i =
  \begin{cases}
    +1 & \text{if} \quad x_i \colon 0\rightarrow 1 \\
    -1 & \text{if} \quad x_i\colon 1\rightarrow 0
  \end{cases}
  ,\quad \forall i\in\{1, \dots, n\}.
\label{flip direction}
\end{equation}
By using $p_i$ and $w_{ik}$, we calculate the efficiency $\bm e$ as
\begin{equation}
    e_i = \alpha \overline{p}_i + (1-\alpha) \sum_k \beta_k \overline{w}_{ik}, \quad \forall i\in\{1, \dots, n\}.
\label{efficinecy}
\end{equation}
Here, $\overline{p}_i=(-p_i)/{\max_{i}(-p_i)}$ and $\overline{w}_{ik}=(-w_{ik})/{\max_{i}(-w_{ik})}$, which are the normalizations of $p_i$ and $w_{i,k}$, respectively. $\alpha$ is the hyperparameter that controls the trade-off between the contributions of the objective function and the constraint satisfaction. The way to set the weight $\bm \beta$ will be mentioned later. 

Next, we describe the specific procedure of the two processes. The feasibility restoration process begins with an infeasible initial solution ${\bm x}_{\rm init} \in\{0,1\}^{n}$. We construct ${\bm x}_{\rm init}$ from the continuous relaxation solution $X_{ii}=\sum_{p\in\overline{\mathcal{P}}}\lambda_px_i^p$ as follows: 
\begin{equation}
  x^{\rm init}_i=
  \begin{cases}
    1 & \text{if}\quad \sqrt{X_{ii}}> 0.5 \\
    0 & \text{otherwise}
  \end{cases}
  ,\quad \forall i\in\{1, \dots, n\}.
\label{rounding}
\end{equation}
In the feasibility restoration process, the weight $\bm \beta$ is defined as $\beta_k=v_k / \sum_{k}v_k$, where $v_k=\max\{0,\sum_{ij}A_{kij}x_i x_j - b_k\}$ is the degree of constraint violation for the constraint $k$ ($\bm x$ is the tentative solution). Using the efficiency $\bm e$, we iteratively flip variables to reduce constraint violations. At each step, the variable with the highest efficiency is flipped, and efficiency is recalculated. This process is repeated until a feasible solution is obtained. 
However, when the constraints are particularly strict or the initial solution is poor, a feasible solution may not be found within a reasonable number of iterations. Therefore, we impose an upper limit $T$ on the number of flips in the feasibility restoration process. If no feasible solution is obtained after $T$ flips, the algorithm terminates without returning a solution.

Subsequently, the local optimization process seeks to improve the objective value while maintaining feasibility. Flipping is limited to variables where the objective function can be improved ($p_i<0$), and flips are only accepted if they do not violate any constraints. The weight $\bm \beta$ is defined as $\beta_k = - r_k / \sum_{k}r_k$ with the margin $r_k=b_k-\sum_{ij}A_{kij}x_i x_j$ for each constraint. The overall algorithm is summarized as Alg. \ref{alg:overall_procedure}.

The efficiency concept, used in greedy methods for quadratic knapsack problems (QKPs)\cite{BILLIONNET1996310, 10589643}, is the basis of our postprocessing approach. In the QKP, the matrix $Q$ is an upper-triangular matrix with non-negative elements and $A$ is a diagonal matrix with non-negative elements (with the number of constraints $m = 1$). In this case, because $p_i$ and $w_i$ share the same sign when $x_i$ is flipped, defining the efficiency as the ratio $e_i = p_i / w_i$ allows us to capture how effectively both the objective function value and the satisfaction of the constraint improve. However, for a general problem \eqref{constrained quadratic programming}, both $Q$ and $A_k$ can take positive or negative values, making it impossible to represent efficiency simply as a ratio. Hence, in this study, we define efficiency as in Eq. \eqref{efficinecy} by normalizing $p_i$ and the constraint term $w_{ik}$ and then summing them.

In addition, in a related paper, a postprocessing procedure that remaps infeasible solutions, obtained through quantum computing, to feasible solutions \cite{10472069} has been proposed. This procedure defines a quantity analogous to efficiency by summing the changes in the objective function and constraint terms, and then performs local search on the basis of that quantity. However, since it is designed for the case in which the variables within each constraint are independent, it cannot be generally applied to the broader class of problems \eqref{constrained quadratic programming} targeted in our research. Consequently, our approach can be applied to a wider range of problem settings.

To illustrate its features more clearly, we compare it with a typical Markov-chain-Monte-Carlo (MCMC)-based approach. In MCMC, one Monte-Carlo step examines each spin in turn as a flip candidate, deciding whether to accept or reject the flip based on the local energy difference. Consequently, multiple variables may be flipped in a single step. In contrast, our proposed method computes the energy change (or ``efficiency'') for flipping each variable and then flips only one single variable that offers the greatest improvement. This purely deterministic procedure flips only one variable per iteration and does not incorporate a temperature-based probability of accepting uphill moves.

As a result, our method can be seen as ``rounding-based'' local search that emphasizes the fast computation of a feasible solution, rather than relying on thermal fluctuations to escape local minima. This characteristic makes it possible to rapidly obtain a binary solution that satisfies the constraints and provides a reasonable local optimum.

\begin{algorithm}
\caption{Overall postprocessing Algorithm}
\label{alg:overall_procedure}
\begin{algorithmic}[1]
\Require $Q$, $A$, $b$, initial solution $\bm{x}\in\{0,1\}^N$, parameter $(\alpha_f, \alpha_l),T$
\Ensure Final (local optimum) solution $\bm{x}$

\State \textbf{function}\ \textsc{FeasibilityRestoration }($Q, A, b, \bm{x}, \alpha,T$):
\State \quad \textbf{while} $\bm{x}$ is infeasible \textbf{and} iterations $< T$:
\State \quad \quad Compute $\bm{e}$
\State \quad \quad \textbf{for} $i$ in descending order of $e_i$:
\State \quad \quad \quad \textbf{if} solution with flipped $x_i$ is unexplored:
\State \quad \quad \quad \quad Flip $x_i$
\State \quad \quad \quad \quad \textbf{break for}
\State \quad \textbf{return} $\bm x$

\State \textbf{function}\ \textsc{LocalOptimization }($Q, A, b, \bm{x}, \alpha$):
\State \quad \textbf{repeat}:
\State \quad \quad Compute $\bm{e}$ for variables $i$ with $p_i < 0$
\State \quad \quad Flip $x_i$ with the highest $e_i$
\State \quad \textbf{until} no improvement occurs
\State \quad \textbf{return} $\bm x$

\State $\bm{x}_f \leftarrow$ \textsc{FeasibilityRestoration }($Q,A,b,\bm{x},\alpha_f, T$)
\State \textbf{if} $\bm{x}_f$ is feasible:
\State \quad $\bm{x}_l \leftarrow$ \textsc{LocalOptimization }($Q,A,b,\bm{x}_f,\alpha_l$)
\State \quad \textbf{return} $\bm{x}_l$
\State \textbf{else}:
\State \quad\textbf{return} no feasible solution

\end{algorithmic}
\end{algorithm}

\section{Results}
In this section, we evaluate the solution quality and computation time of the proposed method, which combines CG with postprocessing (hereafter, referred to as CG+pp). To solve the pricing problems, we employ QA using D-Wave Advantage 6.4. The initial solution provided to CG is the trivial feasible solution ${\bm x}_0=(1,0, 0,\dots,0)$, and we begin CG with $\mathcal{P}_0=\{{\bm x}_0\}$. In all experiments, the parameter $\alpha$ in the efficiency \eqref{efficinecy} is set to $\alpha_f=0.1$ for the feasibility restoration process and $\alpha_l=0.9$ for the local optimization process. In all experiments, the maximum number of flips in the feasibility restoration process was fixed at $T=1000$.

The benchmark problems used in this study are the same as those in the prior work\cite{Hirama2023}. Specifically, the elements $Q_{ij}$ and $ A_{kij}$ of the matrices $Q$ and $A_k$ ($1\leq i \leq j \leq n$) are randomly chosen from $\{+1, -1\}$, and the constraint bounds $b_k$ are set to $1$. 

\begin{figure}[tbh]
\centering
\includegraphics[scale=0.7]{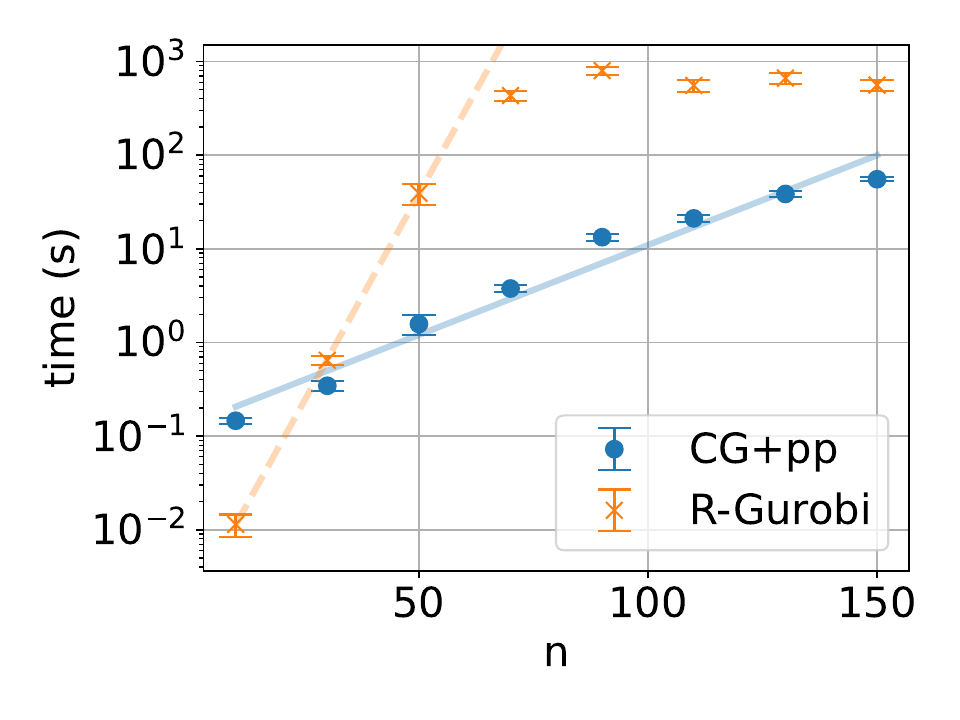}
\caption{(Color online) $n$ dependence of the computation time of CG+pp and R-Gurobi. Exponential fitting curves $f(x)=\exp(ax+b)$ are applied to the data. The fitting parameters are $a=0.04$ and $b=-2.02$ for CG+pp, and $a=0.20$ and $b=-6.52$ for R-Gurobi.}
\label{fig:vs_R_Gurobi}
\end{figure}

First, we compare the computation time of CG+pp with the general-purpose optimization solver Gurobi Optimizer. Gurobi terminates its computation when it reaches an objective function value $\sum_{ij}Q_{ij}x_i x_j$ equivalent to that obtained by CG+pp. This approach is referred to as R-Gurobi. The maximum computation time for R-Gurobi is set to 1000 s, and Gurobi version 11.0.3 is used. CG and the postprocessing method are implemented in Python, and experiments are conducted on a CPU-based system. Figure \ref{fig:vs_R_Gurobi} illustrates the dependence of computation time on the problem size $n$ for CG+pp and R-Gurobi when $m/n=0.2$. In this experiment, CG+pp produced a feasible solution for every instance tested. The plot represents the average computation time for 20 problem instances, and error bars indicate standard errors.

In Fig. \ref{fig:vs_R_Gurobi}, the scaling of the computation time is evaluated using a fitting function $f(x)=\exp(ax+b)$. For CG+pp, the fitting result shows $a=0.04$, wheras for R-Gurobi, $a=0.2$. This indicates that the computation time for CG+pp increases more gradually with $n$ than that for R-Gurobi. This indicates that CG+pp provides a shorter computation time for achieving comparable approximation accuracy when the $m/n$ ratio is small.

Note that for R-Gurobi, a computation time limit of 1000 s was imposed. Consequently, the fitting for R-Gurobi was performed using only the first three data points, where the computation time did not exceed the limit. This constraint emphasizes the rapid increase in computation time for R-Gurobi compared with CG+pp.

\begin{figure}[tbh]
\centering
\includegraphics[scale=0.7]{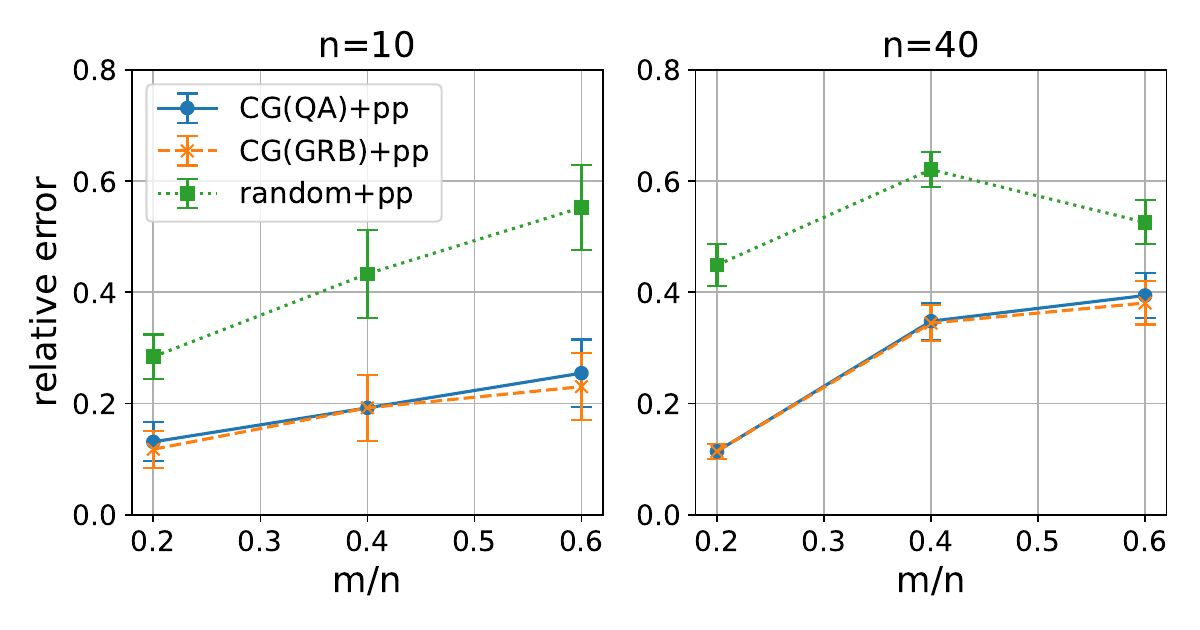}
\caption{(Color online) Dependence of relative error on the $m/n$ ratio for CG(QA)+pp, CG(GRB)+pp, and random+pp when $n=10$ (left panel) and $n=40$ (right panel).}
\label{fig:relative_error}
\end{figure}

In the above experiment, we fixed the $m/n$ ratio to $0.2$. Our empirical evidence suggests that the performance of CG depends on this ratio. Thus, we next examine the solution accuracy of CG+pp as a function of the $m/n$ ratio. For comparison, we also evaluate a method where postprocessing is applied to random solutions, referred to as random+pp. Since random solutions are typically infeasible, both processes (feasibility restoration and local optimization) are applied. 
We also compare the solver QA and the exact solver (Gurobi) for the pricing problem, which we refer to as CG(QA) and CG(GRB), respectively. In CG(GRB), to avoid duplicate solutions, the following constraint is added 
to the pricing problem on the basis of the tentative set of 
extreme points $\overline{\mathcal{P}}$ in each iteration: 
$\sum_{i\in N_0^p} (1-x_i^p ) + \sum_{i\in N_1^p} x_i^p \le N-1,\ \forall p \in \overline{\mathcal{P}},$ where $N_0^p$ denotes the set of variables that take the value 1 in the extreme point $\bm{x}^p$ and $N_1^p$ denotes the set of variables that take the value 0 in $\bm{x}^p$. Solution accuracy is evaluated using the relative error $|(E-E^{\ast})/E^{\ast}|$, where $E^{\ast}$ represents the exact objective function value obtained by Gurobi Optimizer. Figure \ref{fig:relative_error} shows the dependence of relative error on the $m/n$ ratio for $n=10$ and $n=40$. In the second experiment as well, CG(QA)+pp and CG(GRB)+pp succeeded in finding feasible solutions for all instances. The plot represents the average computation time for 50 problem instances and error bars indicate standard errors.

From Fig. \ref{fig:relative_error}, it can be seen that CG+pp consistently achieves lower relative errors than random+pp, regardless of the $m/n$ ratio or problem size $n$. However, the accuracy of CG+pp deteriorates as $n$ and the $m/n$ ratio increase. Indeed, the accuracies of CG(QA) and CG(GRB) are nearly identical.

\begin{figure}[tbh]
\centering
\includegraphics[scale=0.7]{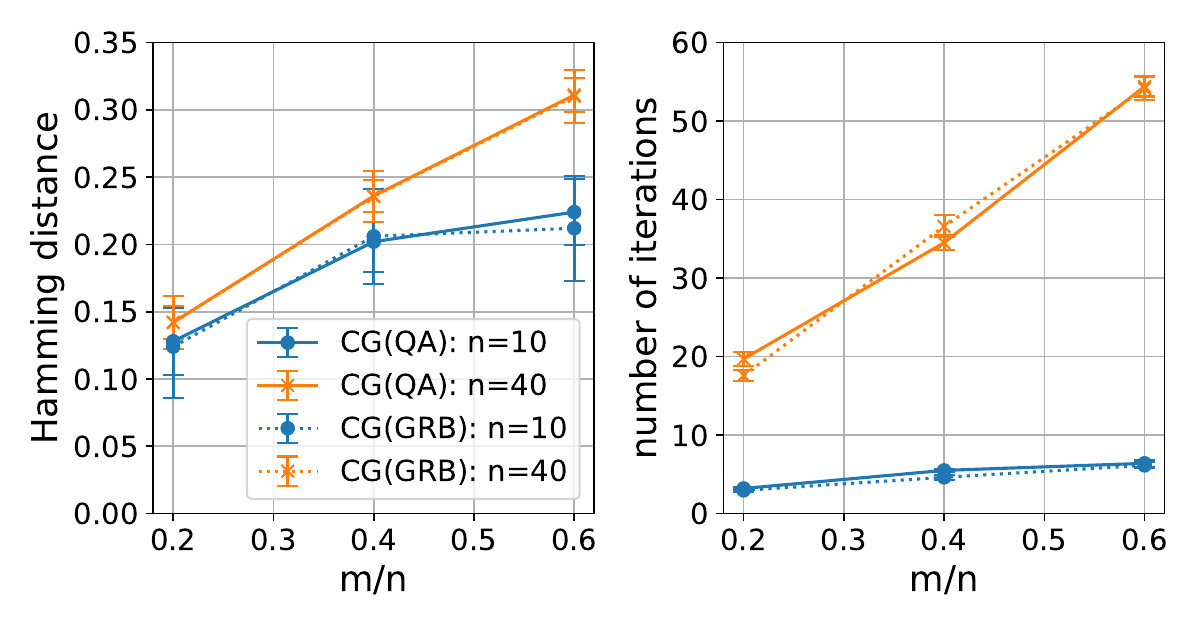}
\caption{(Color online) (Left panel) Dependence of Hamming distance on the $m/n$ ratio for CG(QA) and CG(GRB). (Right panel) Dependence of the number of iterations on the $m/n$ ratio for CG(QA) and CG(GRB).}
\label{fig:ham_dis_num_iter}
\end{figure}

Next, we investigate the accuracy of the rounded solutions obtained from CG \eqref{rounding} as a function of the $m/n$ ratio. The accuracy is measured on the basis of the Hamming distance $\sum_{i=1}^{n}|x_i - x^{\ast}_i|/n$ from the exact solution ${\bm x}^{\ast}$, obtained by Gurobi Optimizer. In addition, we examine the number of iterations required for CG to terminate, which corresponds to the number of added extreme points. Figure \ref{fig:ham_dis_num_iter} shows the results.

From the left panel of Fig. \ref{fig:ham_dis_num_iter}, it is evident that the accuracy of CG(QA) and CG(GRB) deteriorates as $n$ and the $m/n$ ratio increase. The right panel of Fig. \ref{fig:ham_dis_num_iter} shows that the number of iterations required for CG(QA) and CG(GRB) increases with $n$ and the $m/n$ ratio, indicating greater difficulty in solving the problem. Moreover, in all results, the differences between CG(QA) and CG(GRB) are minimal, indicating that there are no significant differences arising from the choice of the solver for the pricing problem.

\begin{figure}[tbh]
\centering
\includegraphics[scale=0.7]{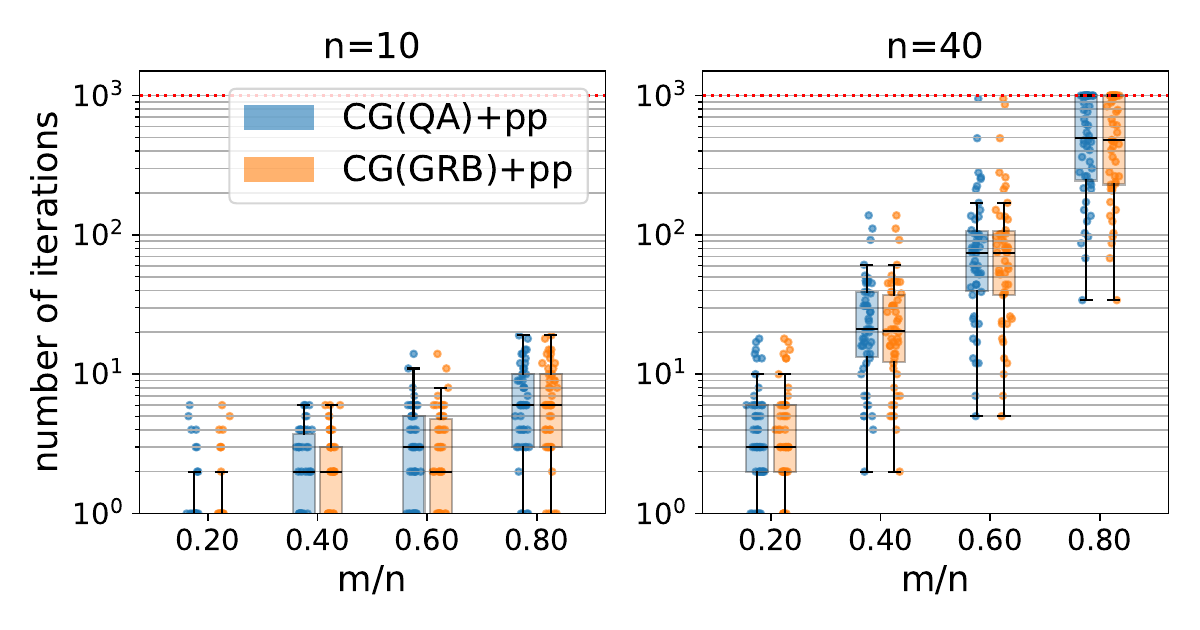}
\caption{(Color online) Dependence of the number of iterations of the feasible restoration process on the $m/n$ ratio for CG(QA)+pp and  CG(GRB)+pp when $n=10$ (left panel) and $n=40$ (right panel). The red dotted line corresponds to $T=1000$.}
\label{fig:num_iter_feasible_restoration}
\end{figure}

We next investigate the number of flips required in the feasibility restoration process and the associated feasibility rate. Figure\ref{fig:num_iter_feasible_restoration} shows how the iteration count of CG(QA)+pp and CG(GRB)+pp depends on the $m/n$ ratio for $n=10$ and $n=40$. We use 50 instances for each setting.

Figure \ref{fig:num_iter_feasible_restoration} indicates that, for $n=10$, the iteration count grows only modestly as the $m/n$ ratio increases. In contrast, for $n=40$, it increases sharply, and at $m/n=0.8$, some instances reach the limit of $T=1000$. In these cases, no feasible solution is produced; 15 of such instances occur for CG(QA)+pp and 14 for CG(GRB)+pp. Consequently, the feasibility rates at $m/n=0.8$ and $n=40$ are 0.72 and 0.70, respectively, showing only a minor difference between the two solvers. Since feasibility depends on the parameters $(\alpha_f, \alpha_l)$ and $T$, tuning them could improve the success rate, although our method still tends to fail when the $m/n$ ratio is large.

\section{Discussion}
In this paper, we proposed a method of deriving binary feasible solutions from the continuous relaxation solutions obtained through CG. As shown in Fig. \ref{fig:vs_R_Gurobi}, the proposed method achieves a shorter computation time for obtaining comparable solution accuracy as the problem size increases, than the general-purpose commercial solver Gurobi Optimizer. However, as illustrated in the left panel of Fig. \ref{fig:ham_dis_num_iter}, the accuracy of solutions obtained by CG deteriorates as the $m/n$ ratio increases, and as shown in the right panel of  Fig. \ref{fig:ham_dis_num_iter}, the number of iterations required for CG also increases. Since the postprocessing method relies on local search starting from an initial solution \eqref{rounding}, its performance is strongly affected by the characteristics of the solutions provided by CG. CG+pp's performance depends on CG, excelling in problems with a small $m/n$ ratio. Moreover, the feasibility restoration process does not guarantee a feasible solution, as shown in Fig. \ref{fig:num_iter_feasible_restoration}. When there are many constraints and it becomes difficult to find a feasible solution solely through local search from the initial solution, the method may struggle to terminate within a practical amount of time. Hence, there remains room for further refinements aimed at enhancing feasibility.

We also compared the approximate solver QA and the exact solver Gurobi for solving the pricing problem. In our experiments, no significant differences were observed between the two regarding solution accuracy and the number of iterations required for CG. This result indicates that the optimality of the solutions for the pricing problems has a limited impact on the final output of CG. In addition, CG(QA) and CG(GRB) were compared for problem sizes $N \le 40$ such that Gurobi could solve the pricing problems in a realistic amount of time. Within this range, there are no substantial differences in computation times between CG(QA) and CG(GRB). However, once the problem size increases beyond this range, the time Gurobi requires to solve the pricing problem increases markedly. Hence, as the problem size increases, CG(QA) is expected to maintain a computation time advantage over CG(GRB) as well. However, further investigation is required to determine the approximation accuracy required to ensure that CG terminates within a realistic number of iterations. 

Another potential avenue for future work is to extend the proposed method to problems involving both equality and inequality constraints. Specifically, the method could be modified to address general quadratic programming problems that include additional equality constraints of the form $\sum_{ij}C_{lij}x_i x_j=d_l,\ \forall l$. Real-world problems often involve both equality constraints, such as one-hot constraints, and inequality constraints, such as capacity constraints. Expanding the proposed method to handle such problems would enhance its applicability to practical scenarios.

{\it Acknowledgement.}
This paper is based on results obtained from a project, JPNP23003, commissioned by the New Energy and Industrial Technology Development Organization (NEDO).

{\it Author contributions.}
T.T. conceived of the presented idea, developed the methodology, and performed the experiments. N.M. managed and supervised the project. T.Y. contributed to the development of the methodology through discussions. R.H. helped to supervise the project. M.O. reviewed the draft. All authors discussed the results and contributed to the final manuscript.

* maruyama@sigmailab.com
\bibliographystyle{jpsj}
\bibliography{18102FP}
\end{document}